\documentclass[aps,prl,twocolumn,superscriptaddress,floatfix]{revtex4-2}

\usepackage{graphicx}
\usepackage{dcolumn}
\usepackage{bm}
\usepackage{empheq}
\usepackage[outdir=./]{epstopdf}
\usepackage{verbatim}
\usepackage{amsmath}
\usepackage{mathrsfs}
\usepackage{mathtools}

\newcommand{\sig}[1]{\sigma_{\mathbf{#1}}}
\usepackage{bm}
\usepackage[section]{placeins}

\begin{document}

\title{Elastic Forces Drive Nonequilibrium Pattern Formation in a Model of Nanocrystal Ion Exchange}

\author{Layne B. Frechette}
\affiliation{Department of Chemistry, University of California, Berkeley, California 94720, USA}
\affiliation{Erwin Schr{\"o}dinger Institute for Mathematics and Physics, University of Vienna, Boltzmanngasse 9, Wien 1090, Austria}
\author{Christoph Dellago}
\email{christoph.dellago@univie.ac.at }
\affiliation{Erwin Schr{\"o}dinger Institute for Mathematics and Physics, University of Vienna, Boltzmanngasse 9, Wien 1090, Austria}
\affiliation{Faculty of Physics, University of Vienna, Boltzmanngasse 5, Wien 1090, Austria}
\author{Phillip L. Geissler}
\email{geissler@berkeley.edu}
\affiliation{Department of Chemistry, University of California, Berkeley, California 94720, USA}
\affiliation{Erwin Schr{\"o}dinger Institute for Mathematics and Physics, University of Vienna, Boltzmanngasse 9, Wien 1090, Austria}

\begin{abstract}
 Chemical transformations, such as ion exchange, are commonly employed to modify nanocrystal compositions. Yet the mechanisms of these transformations, which often operate far from equilibrium and entail mixing diverse chemical species, remain poorly understood. Here, we explore an idealized model for ion exchange in which a chemical potential drives compositional defects to accumulate at a crystal’s surface. These impurities subsequently diffuse inward. We find that the nature of interactions between sites in a compositionally impure crystal strongly impacts exchange trajectories. In particular, elastic deformations which accompany lattice-mismatched species promote spatially modulated patterns in the composition. These same patterns can be produced at equilibrium in core/shell nanocrystals, whose structure mimics transient motifs observed in nonequilibrium trajectories. Moreover, the core of such nanocrystals undergoes a phase transition – from modulated to unstructured – as the thickness or stiffness of the shell is decreased. Our results help explain the varied patterns observed in heterostructured nanocrystals produced by ion exchange and suggest principles for the rational design of compositionally-patterned nanomaterials.
\end{abstract}

\maketitle

Methods of chemical transformation are routinely used to alter the properties of nanocrystals post-synthesis \cite{han_interplay_2021}. Among these, ion exchange -- the replacement of one ion species by another in a crystal \cite{son_cation_2004,saruyama_transformations_2021} -- is a common and effective technique for modifying nanoparticle composition. Cation exchange in particular has been used to produce a variety of heterostructured nanocrystals, chiefly metal chalcogenides, whose mixed compositions exhibit diverse morphologies \cite{de_trizio_forging_2016,groeneveld_tailoring_2013}. Despite their utility, the mechanisms which govern the progress of cation exchange reactions remain unclear. Thus, cation exchange has likely not attained its full potential as a method for precisely tuning the composition and spatial organization of species within nanocrystals.  

Experiments probing cation exchange reactions have yielded several key observations. Kinetic measurements of exchange among certain cation species suggest distinct behaviors on different timescales: rapid change in the composition at short times, presumably due to the fast introduction of guest ions at the nanocrystal surface, followed by much slower change at long times, thought to reflect internal diffusion of cations \cite{groeneveld_tailoring_2013,bothe_solid-state_2015,kershaw_investigation_2017,moser_situ_2017}. Meanwhile, product morphologies vary widely, depending on the chemical identity of the species involved and on reaction conditions like stoichiometry \cite{de_trizio_forging_2016,steimle_experimental_2020,saruyama_transformations_2021}. In cases where exchange is only partial, the resulting heterostructures can depend strongly on the relative sizes of the cations. Exchange of species with significant lattice mismatch tends to produce graded, alloy-like structures (as in Zn/CdSe) \cite{groeneveld_tailoring_2013} or spatially modulated patterns with length scales of several lattice spacings (as in Ag$_2$/CdS) \cite{robinson_spontaneous_2007,demchenko_formation_2008,liu_nanointerface_2020}, while similarly-sized ions instead adopt core/shell or Janus morphologies \cite{de_trizio_forging_2016}. However, the limited spatial and temporal resolution of current experimental techniques has frustrated attempts to build a conceptual framework with which to understand these intriguing behaviors. For example, it is still not entirely clear whether ion exchange proceeds primarily via vacancies or interstitial defects \cite{lesnyak_cu_2015,bothe_solid-state_2015,justo_less_2014}. More broadly, the relative contributions of thermodynamics and kinetics in shaping the (potentially metastable) final products of cation exchange have yet to be elucidated in detail.

 In contrast to experiments, computer simulations can fully resolve atomic motions, but are limited to short time and length scales (especially for chemically detailed models.) Two previous simulation studies, which rendered specific materials with a fairly high level of detail, have yielded some interesting mechanistic insights into cation exchange. Ott et al. \cite{ott_microscopic_2014} performed kinetic Monte Carlo simulations of CdS$\rightarrow$Ag$_2$S exchange, parametrized based on density functional theory calculations of defect energies in the bulk crystal. They found evidence for cooperativity among charged defects and identified this as a key driver of exchange. Fan et al. instead performed molecular dynamics simulations of PbS$\rightarrow$CdS exchange, employing ``pseudoligands'' as a coarse-grained representation of solvent and ligands which typically decorate nanocrystals. They found that pseudoligands promote extraction of Pb and that interstitial defects mediate internal impurity transport, at least at the elevated temperatures necessary to observe significant exchange in the simulations. However, in both cases the authors were only able to monitor reactions on time scales far short of complete exchange. Moreover, neither study explicitly addressed lattice mismatch, which can produce long-ranged elastic interactions \cite{frechette_consequences_2019,frechette_origin_2020,demchenko_formation_2008}.
 
  \begin{figure*}
    \centering
    \includegraphics[width=0.75\linewidth]{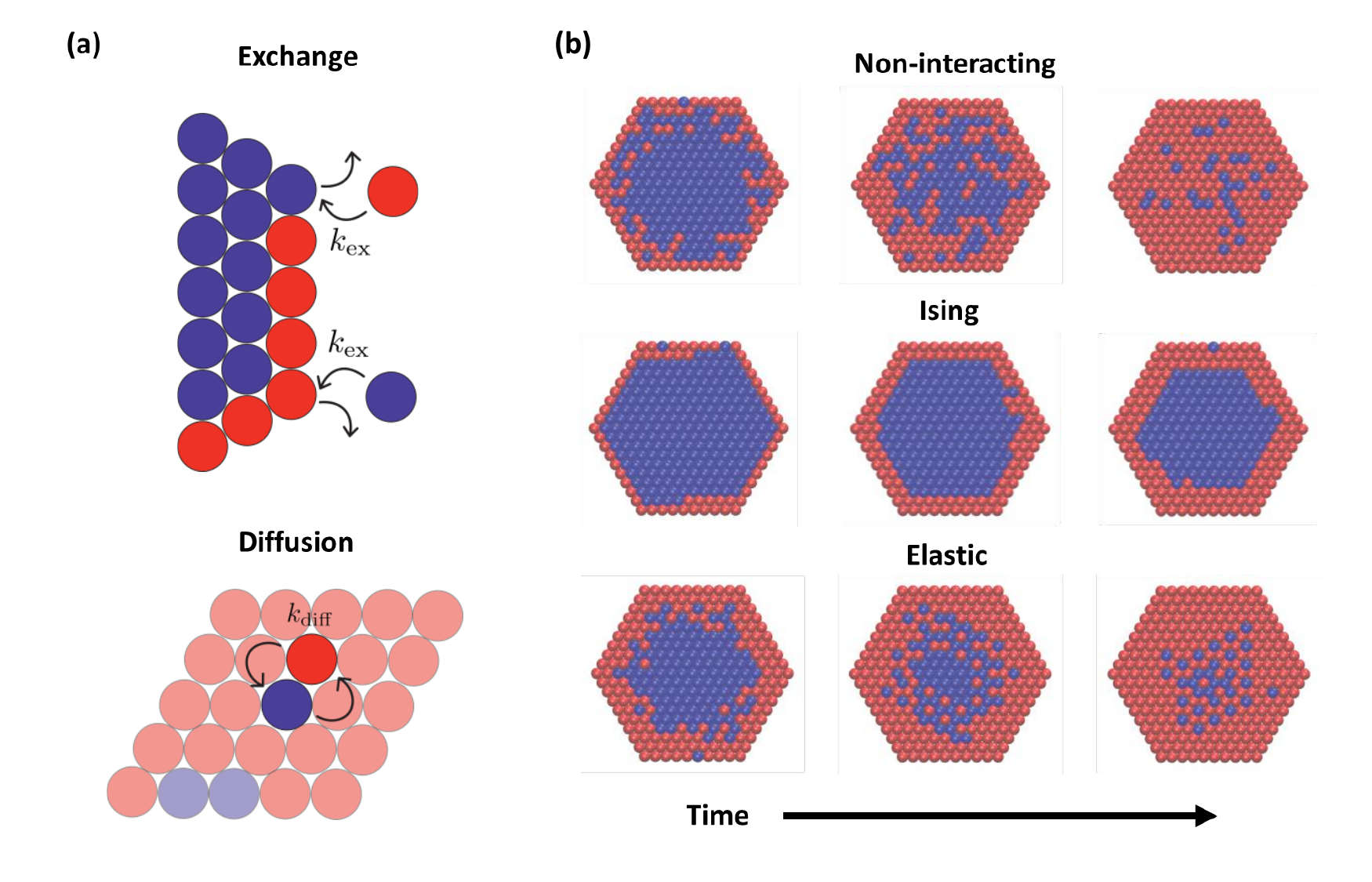}
    \caption{\textbf{(a)} Schematics illustrating exchange at the surface of a nanocrystal (top) and diffusion within a nanocrystal (bottom). Note that $k_{\text{ex}}$ and $k_{\text{diff}}$ depend on the compositional state of the nanocrystal before and after the corresponding event; the value of $k_{\text{ex}}$ is therefore different for the two surface exchange events shown. \textbf{(b)} Representative configurations taken along KMC exchange trajectories for a statically uncorrelated model (Non-interacting), a model with nearest-neighbor interactions (Ising), and a model with the nonlocal, mechanically mediated interactions of Eq. \ref{eq:elastic} (Elastic).}
    \label{fig:figure1}
\end{figure*}

Here, we report computer simulations of a lattice model for ion exchange in nanocrystals. We have in mind cation exchange reactions of metal chalcogenides, but the model does not specify a particular compound. Instead we consider a minimal model which focuses on the key experimental features outlined above: differing timescales for surface exchange and internal diffusion, and lattice mismatch. Although our approach sacrifices a significant degree of microscopic chemical detail, the resulting reduction in computational expense carries benefits that are unprecedented in numerical studies of ion exchange. First, we are able to establish equilibrium properties of the microscopic model not just in reactant and product states, but also across the entire range of intermediate compositions, providing a key reference for understanding reaction dynamics. Second, we can propagate exchange trajectories all the way from reactants to products, without heating the system unrealistically or artificially biasing its evolution. Finally, we can thoroughly sample ensembles of nonequilibrium exchange trajectories, enabling conclusions and comparisons that are statistically robust. 

\section*{Model}

Our model resolves compositional change on a scale that is microscopic but not atomic. We associate a binary variable $\sigma_{\mathbf{R}}=\mp 1$ with each site $\mathbf{R}$ of a finite lattice, roughly equivalent to dividing a nanocrystal into $N$ unit cells that have either exchanged or not. $\sigma_{\mathbf{R}}= -1$ thus indicates local enrichment of the incoming cation (labeled B and colored red in figures), while $\sigma_{\mathbf{R}}= +1$ indicates enrichment of the outgoing species (labeled A and colored blue). In this way, we represent crystal structure at a coarse level compared to previous simulation studies.  

Exchange is driven by a difference $\Delta\mu = \mu_{\rm B} - \mu_{\rm A}$ between the chemical potentials of A and B cations, i.e., an effective energy $-(\Delta \mu/2)\sum_{\bf R}\sigma_{\mathbf{R}}$. This thermodynamic bias could represent a difference between ion concentrations in a nanocrystal's solution-phase environment, a difference in their solubilities (possibly mediated by dissolved ligands), and/or a difference in their lattice energies \cite{beberwyck_cation_2013}. In typical experiments, conditions for ion exchange are strongly favorable, so we set the magnitude of $\Delta \mu$ to be much larger than the thermal energy $k_{\rm B}T$.

Trajectories of this model advance through discrete, reversible events in which the state of one or more lattice sites changes. At a site ${\mathbf{R}}_{\text{b}}$ on the nanocrystal's boundary, one cation type can be exchanged for another ($\sigma_{{\mathbf{R}}_{\text{b}}} \rightarrow -\sigma_{\mathbf{R}_{\text{b}}}$), with a rate $k_{\text{ex}}$ that depends on the system's compositional configuration $\{ \sigma_{{\mathbf{R}}}\}$. The nanocrystal's interior evolves through diffusive steps that swap the states of adjacent lattice sites ($\sig{R} \rightarrow \sig{R'}, \sig{R'} \rightarrow \sig{R}$, where ${\mathbf{R}}$ and ${\mathbf{R}}'$ are nearest neighbors), with configuration-dependent rate $k_{\text{diff}}$. Such a swap may represent a series of microscopic barrier-crossing events, perhaps involving transient vacancies or interstitials \cite{lesnyak_cu_2015,bothe_solid-state_2015,justo_less_2014}; we resolve only the net transport of ion density. See \ref{fig:figure1}a for illustrations of these events. The rates $k_{\text{ex}}$ and $k_{\text{diff}}$ are formulated to satisfy detailed balance with respect to the equilibrium probability distribution of the nanocrystal's compositional state. Because they depend on configuration $\{ \sigma_{{\mathbf{R}}}\}$, their values generally change as the reaction proceeds. Similar to previous simulations by Ott et al. \cite{ott_microscopic_2014}, these stochastic rate processes are numerically realized with a kinetic Monte Carlo (KMC) algorithm (see Materials and Methods).

\section*{Ion Exchange Dynamics}

\begin{figure}
    \centering
    \includegraphics{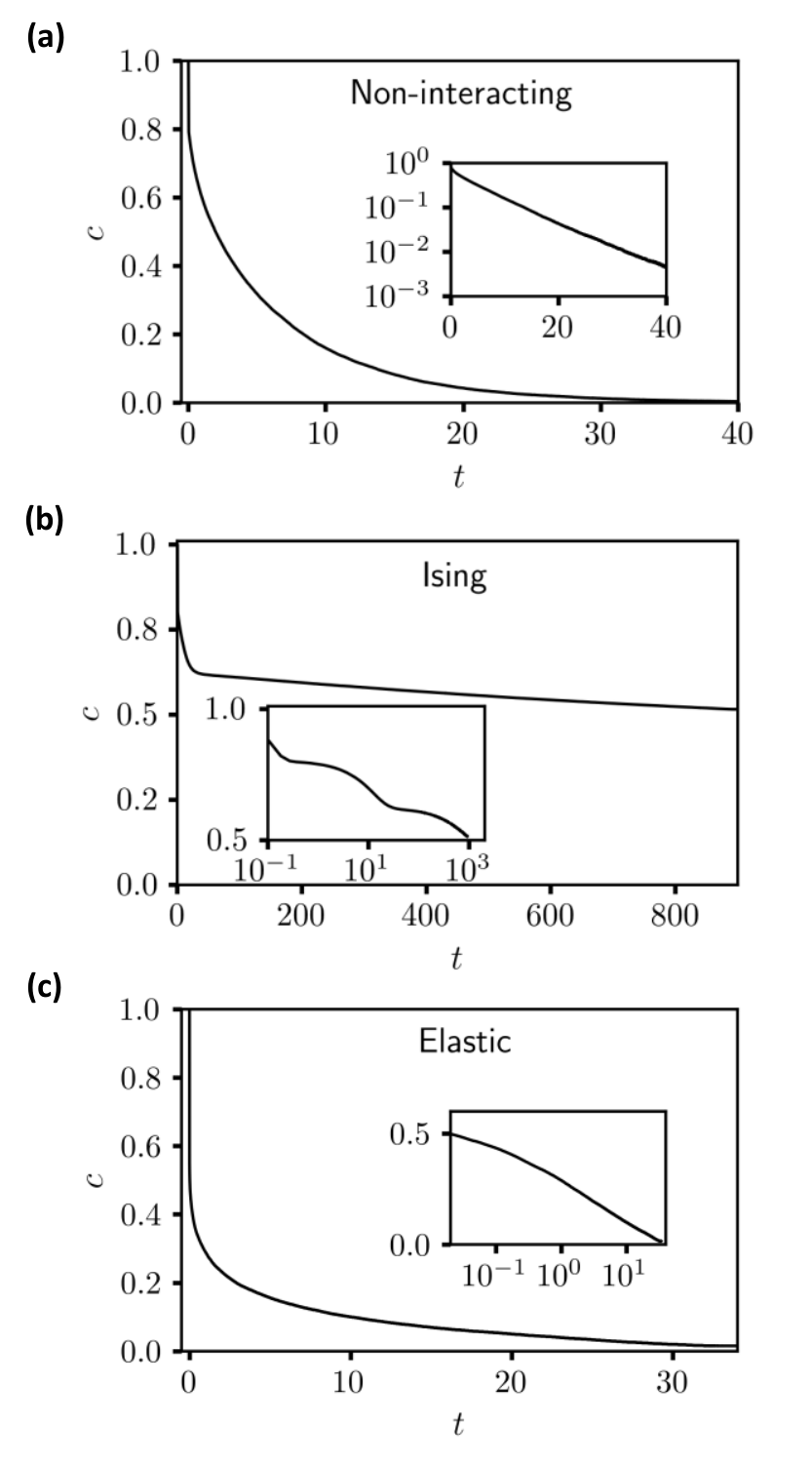}
    \caption{Net composition versus time averaged over 100 KMC trajectories of \textbf{(a)} the non-interacting model with $\Delta \mu/k_BT=-10$, \textbf{(b)} the Ising model with $k_BT/J=2$ and $\Delta \mu/J=-10$, and \textbf{(c)} the elastic model with $k_BT/\epsilon=0.2$ and $\Delta\mu/\epsilon=-20$. In each case, $N=271$. The inset in \textbf{(a)} shows the same data with composition on a logarithmic scale, emphasizing the approximately exponential decay of the composition. The insets in \textbf{(b)} and \textbf{(c)} show the same data with time on a logarithmic scale. The steplike decay of composition apparent in the inset of \textbf{(b)} reflects monolayer-by-monolayer advancement of the B-rich shell into the crystal interior. Modified from Reference \citenum{frechette_chemical_2020}.}
    \label{fig:figure2}
\end{figure}

\begin{figure*}
    \centering
    \includegraphics[width=\linewidth]{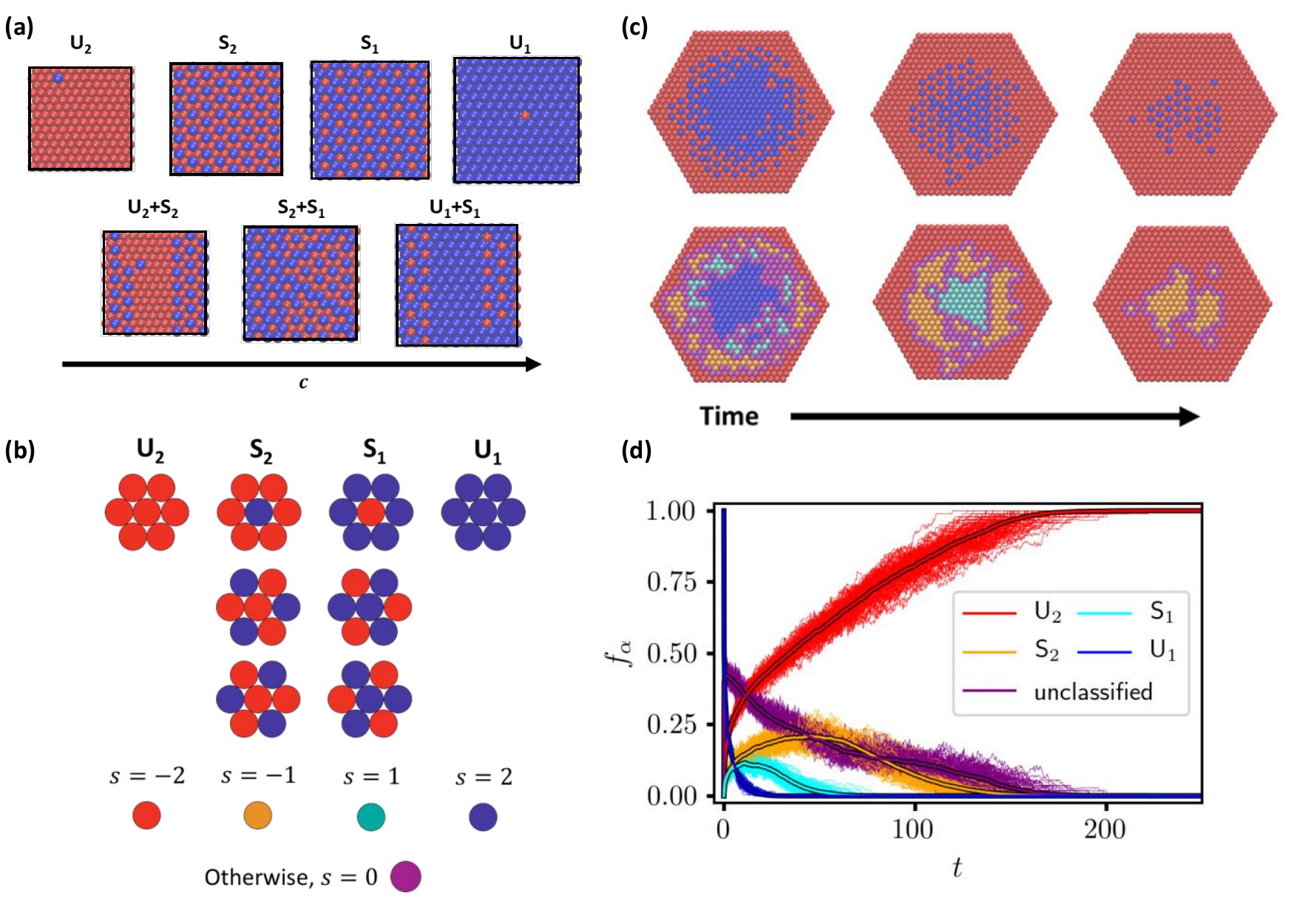}
    \caption{\textbf{(a)} Representative configurations taken from bulk simulations of the elastic model at equilibrium (see Reference \citenum{frechette_consequences_2019}). Across a range of net compositions $c$, unstructured ($U_j$, $j=1,2$) and superlattice ($S_j$, $j=1,2$) phases, as well as coexistence between these phases ($U_j$+$S_j$, $S_1$+$S_2$), can be observed. \textbf{(b)} Seven-site plaquettes associated with different phases, together with corresponding values of the order parameter $s$ and the color assigned to the central site of each plaquette. All other plaquettes are unclassified and assigned $s=0$. \textbf{(c)} Representative configurations along the course of a KMC exchange trajectory of the elastic model shown in the original color scheme (top) and in the plaquette color scheme (bottom). \textbf{(d)} Fractional population $f_{\alpha}$ of each phase $\alpha$ over time in KMC trajectories of the elastic model with $N=721$. Solid-color lines represent individual trajectories; the thick lines with black outlines represent averages over 100 independent trajectories.}
    \label{fig:figure3}
\end{figure*}

If governed only by a difference in chemical potential (and not by configuration-dependent energies), the model exchange dynamics would amount to simple diffusion in a domain bounded by a sink. A trajectory exemplifying this ``non-interacting" case is shown in \ref{fig:figure1}b for a hexagonal lattice in two dimensions. A corresponding plot of the net composition over time is shown in \ref{fig:figure2}a. With $|\Delta \mu|\gg k_{\rm B}T$, A$\rightarrow$B exchange at the perimeter is extremely facile and nearly irreversible. Mirroring experimental observations \cite{groeneveld_tailoring_2013,bothe_solid-state_2015,kershaw_investigation_2017,moser_situ_2017}, the nanocrystal's outermost layer is transformed immediately. Interior swap moves, which do not change the system's net composition $c=(\sum_{\mathbf{R}}\sigma_{\mathbf{R}}/N+1)/2$, leave the energy unchanged in this simple case. Neither facilitated nor hindered, diffusion proceeds slowly, transporting B inward and A outward with a high degree of randomness at this length scale.

Cation exchange dynamics are made interesting by interactions, which correlate the composition of different sites, in space and in time, and can generate surprising collective behaviors. We consider two kinds of site-site interactions that have figured importantly in studies of alloy thermodynamics \cite{weinkamer_using_2004}. One is spatially local,
\begin{equation}
    E_{\text{Ising}}=-J\sum_{\langle\mathbf{R},\mathbf{R}'\rangle}\sigma_{\mathbf{R}}\sigma_{\mathbf{R}'},
\end{equation}
where the sum is restricted to nearest neighbors on the lattice and $J$ is a constant. This Ising-like interaction describes short-ranged chemical preferences for sites of the same type ($J>0$) or of opposite type ($J<0$), and its influence on the time dependence of phase change has been investigated extensively \cite{bortz_time_1974,marro_time_1975,rao_time_1976,weinkamer_using_2004}. The other energetic contribution is mechanical in origin, accounting for the elastic stress inherent to mixing solid materials with different lattice constants \cite{fratzl_ising_1995,fratzl_ising_1996,fratzl_modeling_1999}. We adopt a simple and general description of these elastic forces, in which neighboring sites of the lattice prefer a bond length that depends on the identities of their occupants \cite{frechette_consequences_2019}. Integrating out fluctuations in lattice geometry yields an effective energy for the composition variables
\begin{equation}
    E_{\rm elastic} = \frac{\epsilon}{2}\sum_{{\bf R}, {\bf R}'} \sig{R} V_{\rm elastic}({\bf R},{\bf R}') \sig{R'},\label{eq:elastic}
\end{equation}
where $\epsilon$ is a positive constant. The effective elastic interaction potential $V_{\rm elastic}({\bf R},{\bf R}')$, whose form is detailed in Materials and Methods, decays gradually and non-monotonically with distance. While $E_{\rm elastic}$ is minimum for unmixed states (pure A or pure B), at intermediate compositions it can favor spatial patterns that are richly modulated (see \ref{fig:figure3}a and Reference \citenum{frechette_consequences_2019}). Its consequences for the dynamics of compositional phase change have, to our knowledge, not been thoroughly explored, especially for nanoscale systems. 

An effort to mimic a real material would include both Ising and elastic contributions, but here we consider their impact separately (either $J=0$ or $\epsilon=0$). Our exploration of the combined case (both $J$ and $\epsilon$ nonzero) suggests a similar range of behaviors. Ising interactions, for $J>0$ and well below the critical temperature, greatly impede the overall kinetics of exchange, as illustrated in the trajectory shown in \ref{fig:figure1}b and in \ref{fig:figure2}b. In this case rapid exchange at the perimeter is followed by much slower advance of a sharp A/B interface towards the center of the nanocrystal. Since A species can only be removed from the nanocrystal at its boundary, this process requires that A atoms enter and cross the B-rich shell as high-energy impurities. The costly creation of these defects is rate-limiting at the temperatures of interest, and is especially hindered at smooth regions of the A/B interface where many favorable interactions must be disrupted simultaneously. Experiments on lead chalcogenide nanocrystals in which Pb$^{2+}$ is exchanged for Cd$^{2+}$ show strong kinetic trapping after exchange of the first few monolayers \cite{casavola_anisotropic_2012,justo_less_2014,fan_atomistic_2016,nelson_nanocrystal_2019}, akin to our simulated trajectories and consistent with the bulk immiscibility of PbS and CdS \cite{bethke_sub-solidus_1971}. This exchange scenario is strongly foreshadowed by equilibrium thermodynamics of the Ising model at sub-critical conditions \cite{chandler_introduction_1987}. Whether in bulk or in a nanocrystal, nearly pure phases dominate here, and the line tension of the A/B interface is high, ensuring strong segregation and high barriers to impurity transport.

The relationship between thermodynamics and exchange kinetics is more subtle for the elastic model. In previous work \cite{frechette_consequences_2019,frechette_origin_2020} we have broadly examined the equilibrium behavior of this model (reviewed in \ref{fig:figure3}a), which features microscopically patterned superlattice phases (denoted $S_1$ and $S_2$), in addition to unstructured phases ($U_1$ and $U_2$) with Ising-like symmetry breaking. Stabilities of these phases, and states of coexistence between them, are unusually sensitive to boundary conditions. In particular, the spatially modulated phases $S_1$ and $S_2$ dominate at intermediate composition and low temperature for periodic bulk systems, but are unstable when boundaries are free to deform heterogeneously. At equilibrium the nanocrystals studied in this paper, due to their free boundaries, exhibit superlattice structure with a probability that is negligibly small, vanishing in the limit of large crystal size \cite{frechette_origin_2020,frechette_chemical_2020}. In the nanocrystal's nonequilibrium exchange trajectories, however, superlattice patterns are pronounced. As shown in \ref{fig:figure1}b, motifs of $S_1$ and $S_2$ emerge shortly after the perimeter transforms and dominate the crystal's interior as exchange continues. These trajectories appear not to involve $U_1$/$U_2$ interfaces (i.e., A/B interfaces), and they proceed to completion much more rapidly than for the Ising model (see \ref{fig:figure2}b,c).

In order to quantify transient superlattice patterns in nanocrystal KMC simulations, we introduce an order parameter $s$ that locally distinguishes among the four phases of the elastic model (see Materials and Methods). As illustrated in \ref{fig:figure3}b, the instantaneous value of $s$ at each lattice site $\mathbf{R}$ is determined by the compositional arrangement of the seven-site plaquette comprising $\mathbf{R}$ and its nearest neighbors. The proportion of phase $\alpha=U_1,S_1,...$ can then be judged by the fraction $f_{\alpha}$ of sites with the corresponding value of $s$. This designation also allows a vivid rendering of KMC trajectories, with each site colored according to its current value of $s$ (rather than the value of $\sigma_{\mathbf{R}}$). \ref{fig:figure3}c shows an exchange trajectory visualized in this way, emphasizing the extent and shape of each phase's domain as the reaction proceeds. The clearly evident decay of $U_1$, through $S_1$ and $S_2$, to $U_2$ is demonstrated quantitatively in \ref{fig:figure3}d by plotting $f_{\alpha}$ as a function of time for a large number of exchange trajectories. The proportion of $U_1$ rapidly decays to nearly zero, and the populations of $U_2$, $S_1$, and $S_2$ increase accordingly. A much slower decay of superlattice phases follows, accompanied by terminal growth of $f_{U_2}$ to 1. Though transient, modulated spatial order is thus long-lived during cation exchange of our simulated nanocrystals.

Why do superlattice phases characteristic of the bulk elastic model appear in our out-of-equilibrium nanocrystal, despite their thermodynamic instability under these boundary conditions? We argue that the rapid dynamics of exchange at the perimeter -- the first step in all our exchange trajectories -- produces an effective, transient change in boundary conditions experienced by the crystal's interior. The stiffness of the exchanged shell mimics the influence of a bulk periodic environment, which favors modulated order. An adiabatic exchange dynamics, in which the nanocrystal adjusts completely after each small change in net composition, would begin in a similar fashion, accomodating impurities at the perimeter. But in that reversible case continued accumulation of B atoms would result in not a shell, but instead compact domains of $U_1$ and $U_2$ whose interface spans the nanocrystal. Such shell disintegration is prohibitively slow on the time scale of nonequilibrium exchange at large $\Delta \mu$, so that our irreversible trajectories follow a qualitatively different route. 

\section*{Equilibrium Phases of Core-Shell Nanocrystals}

\begin{figure}
    \centering
    \includegraphics{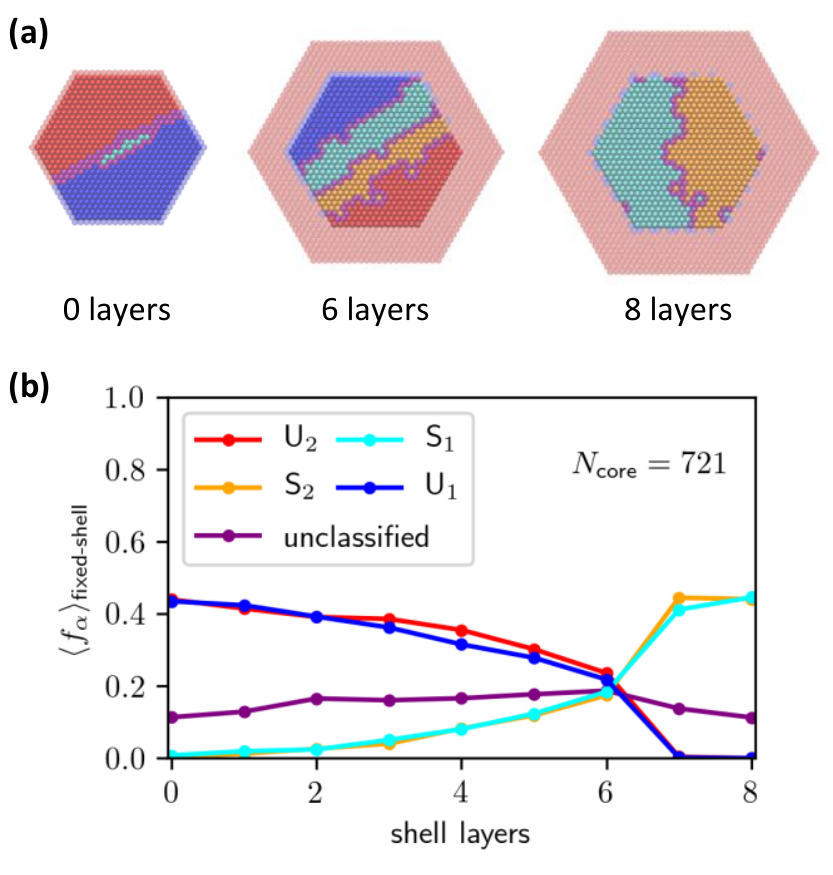}
    \caption{\textbf{(a)} Configurations taken from equilibrium MC simulations of a nanocrystal whose shell composition is fixed at $c=0$, while the core fluctuates with net composition $c=1/2$. Results are shown for $N_{\text{core}}=721$ and for three different shell thicknesses. Interior core sites are colored according to the plaquette scheme; the outermost core sites and the shell sites are translucent and colored according to the original scheme. Note the coexistence between $U_1$ and $U_2$ for 0 shell layers, between $U_1$, $S_1$, $S_2$, and $U_2$ for 6 shell layers, and between $S_1$ and $S_2$ for 8 shell layers. \textbf{(b)} Average fractional population $\langle f_{\alpha}\rangle_{\text{fixed-shell}}$ of the different phases versus shell thickness, measured in atomic layers, for a core size of $N_{\text{core}}=721$. Note the crossover between predominantly unstructured coexistence and predominantly superlattice coexistence that occurs around 6 shell layers.}
    \label{fig:figure4}
\end{figure}

\begin{figure}
    \centering
    \includegraphics{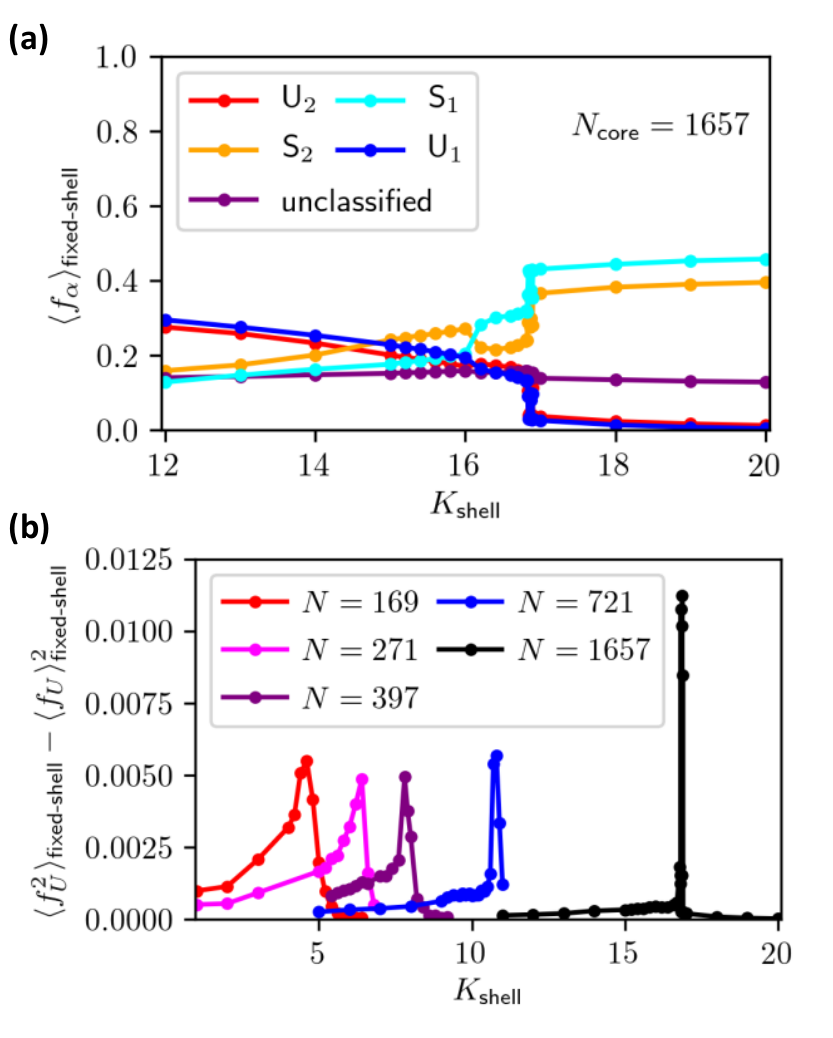}
    \caption{\textbf{(a)} Average fractional population $\langle f_{\alpha} \rangle_{\text{fixed-shell}}$ of the different phases as a function of shell stiffness $K_{\text{shell}}$ for a single-layer shell and core size of $N_{\text{core}}=1657$. Note the sharp change in $\langle f_{\alpha} \rangle_{\text{fixed-shell}}$ occurring at $K_{\text{shell}}\approx 17$, consistent with a first-order phase transition. \textbf{(b)} Variance of the fraction $f_U$ of unstructured phases ($U_1$, $U_2$) versus $K_{\text{shell}}$ for different system sizes (here $N$ refers to the size of the core). Peaks become narrower and higher with increasing system size, consistent with a first-order transition. Peak positions shift to higher $K_{\text{shell}}$ as system size increases due to the decreasing ratio of shell size to core size.}
    \label{fig:figure5}
\end{figure}

We scrutinize this shell hypothesis by using constrained equilibrium Monte Carlo (MC) simulations (see Materials and Methods) to compute equilibrium averages (denoted as $\langle \cdots\rangle_{\rm fixed-shell}$) of $f_{\alpha}$ when the properties of a nanocrystal's shell are systematically varied. We first consider a multi-layer shell consisting of B sites surrounding a core containing $N_{\text{core}}=721$ sites with fixed net composition $c=1/2$ (i.e. half A, half B). A and B atoms may swap positions within the core, but the identity of the shell sites is held fixed. \ref{fig:figure4}a,b shows configurations representative of this constrained equilibrium for several shell sizes, along with $\langle f_{\alpha}\rangle_{\rm fixed-shell}$ in the nanocrystal core as a function of shell thickness. In the absence of a shell, the core is predominantly comprised of $U_1$ and $U_2$ phases. As layers are added to the shell, regions of $S_1$ and $S_2$ gradually become larger and those of $U_1$ and $U_2$ become smaller. At six layers, $f_{\alpha}$ is roughly equal for all $\alpha$. When an additional layer is added, $U_1$ and $U_2$ phases abruptly vanish, and $S_1$ and $S_2$ phases dominate, hinting at a phase transition. Demonstrating the existence of a phase transition in the thermodynamic limit is complicated by our inability to vary the shell thickness continuously in this model. As an alternative, we vary the stiffness of a shell with fixed size, which we choose here to be a single atomic layer. (In practice this is accomplished by changing the spring constant $K_{\text{shell}}$ between bonds connecting shell sites; see Materials and Methods for more details.) The resulting plots of $f_{\alpha}$ versus $K_{\text{shell}}$ (\ref{fig:figure5}a) exhibit the same general behavior, but the transition point can be located with much higher precision. The fraction of $U_1$, $U_2$ drops precipitously (and that of $S_1$, $S_2$ sharply rises) at $K_{\text{shell}}\approx 17$ for $N_{\rm core}=1657$. An underlying phase transition would manifest as a peak in the related fluctuation quantity $\langle f_{U}^2 \rangle_{\rm fixed-shell}-\langle f_U \rangle_{\rm fixed-shell}^2$ (where $f_U=f_{U_1}+f_{U_2}$) whose height grows with system size, diverging in the thermodynamic limit. \ref{fig:figure5}b shows simulation results consistent with this scenario.
The transient superlattices in our nonequilibrium simulations thus appear to be stabilized by a nearby equilibrium phase transition.

\begin{figure}
    \centering
    \includegraphics{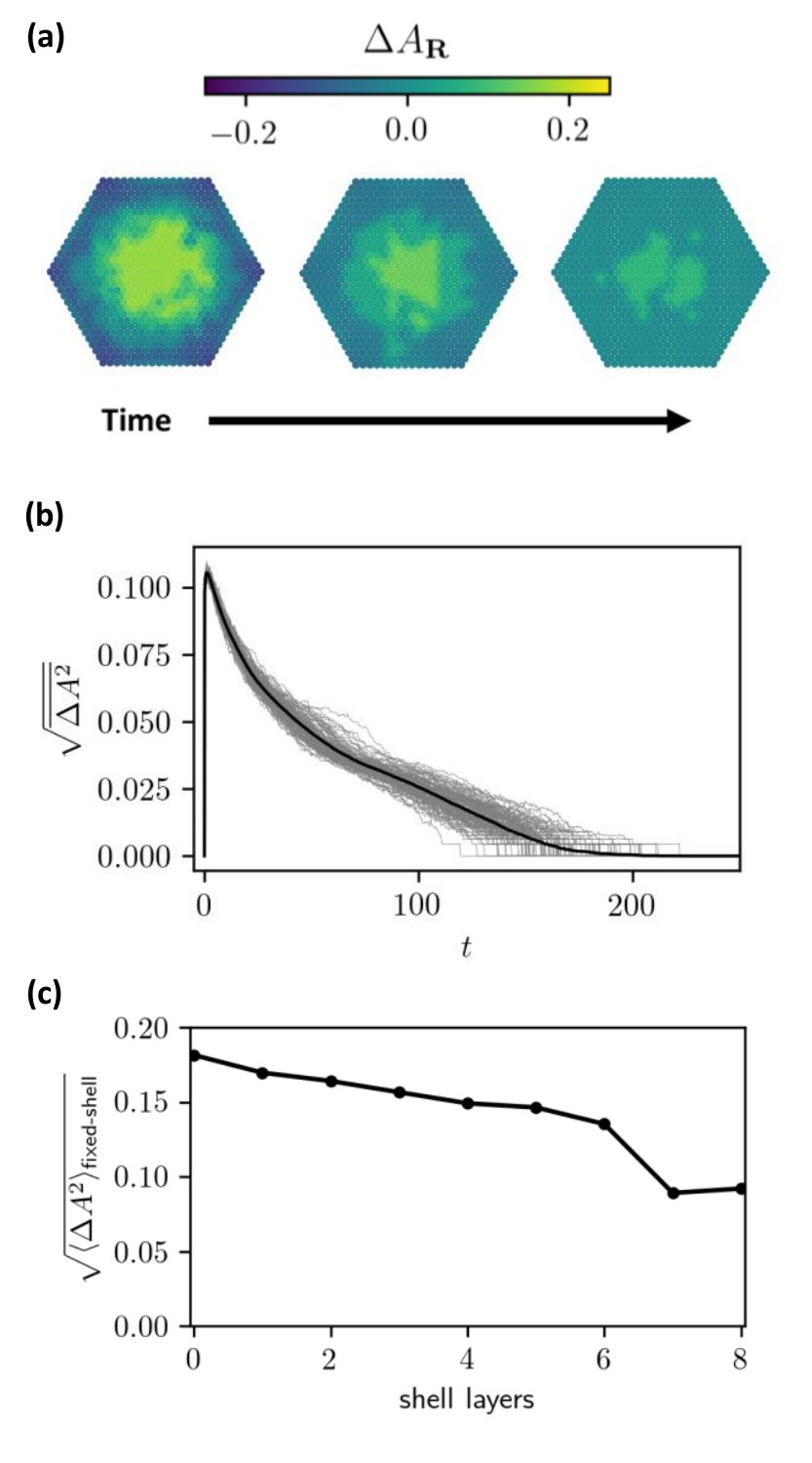}
    \caption{\textbf{(a)} Configurations taken from a KMC exchange trajectory of the elastic model with sites colored according to local area deviation. Compressed regions at the exterior correspond to B-rich regions, while expanded areas in the core correspond to A-rich regions; different shades of color track regions populated by different phases. \textbf{(b)} Root mean square (RMS) deviations of $\Delta A_{\bf R}$ (spatially averaged) versus time. Gray lines represent individual KMC trajectories (with $N=721$), while the black line represents the average over 100 trajectories. \textbf{(c)} RMS deviations (averaged over both space and sampled configurations) from equilibrium MC simulations of fixed-shell nanocrystals (with $N_{\text{core}}=721$) versus shell thickness. Note the drop between 6 and 7 shell layers, corresponding to the transition from a state in which all phases coexist (at 6 layers) to a state of two-phase coexistence involving $S_1$ and $S_2$ only (at 7 layers.)}
    \label{fig:figure6}
\end{figure}

The observed relationship between shell stiffness and modulated order can be explained from an understanding of the stability of superlattice phases in periodic bulk systems at equilibrium. In a fixed volume with regular shape, the coexistence of solid phases with different lattice parameters carries a cost that is extensive in system size (absent topological defects in lattice connectivity) \cite{frechette_consequences_2019,frechette_origin_2020}. The unstructured phases $U_1$ and $U_2$ of the elastic model in 2d have maximally different values of the area per lattice site $A$, so their coexistence is considerably disfavored. The superlattice phases $S_1$ and $S_2$ are more similar to one another in this respect, and hence can dominate at intermediate composition. A nanocrystal with free boundaries can undergo collective elastic distortions which relax this  energetic cost, stabilizing the coexistence of unstructured phases. Adding a large or stiff shell to the nanocrystal hinders such deformations and thus favors superlattice phases.

In \ref{fig:figure6} we view nanocrystal exchange trajectories through this mechanical lens, focusing on the local area per site $A_{\mathbf{R}}$ at each position $\mathbf{R}$ (see Materials and Methods for more details). \ref{fig:figure6}a shows the deviation $\Delta A_{\mathbf{R}} = A_{\mathbf{R}} - \overline{A}$ of this quantity away from its spatial average $\overline{A} = N^{-1}\sum_\mathbf{R} A_{\mathbf{R}}$, for a few configurations taken from a single KMC trajectory; its instantaneous variance $\overline{\Delta A^2} = N^{-1}\sum_\mathbf{R} (\Delta A_{\mathbf{R}})^2$ is plotted in \ref{fig:figure6}b. The spatial distribution is its most heterogeneous at short times: the shell is highly compressed compared to the core, reflecting the dominance of B sites at the perimeter of the crystal and A sites in the interior. As time progresses, this mechanical heterogeneity becomes less dramatic, as reflected in the decay of $\overline{\Delta A^2}$. Additionally, we observe distinct regions with characteristic values of $\Delta A_{\mathbf{R}}$. These regions align precisely with domains of the different compositional phases $\alpha$. The growth of $S_1$ and $S_2$ phases (which have areas intermediate that of $U_1$ and $U_2$) relaxes strain at the core/shell interface, yielding the observed decrease in $\overline{\Delta A^2}$. The corresponding constrained equilibrium average $\langle \Delta {A}^2 \rangle_{\rm fixed-shell} =  N^{-1}\sum_\mathbf{R} \langle(A_{\mathbf{R}} - \langle A\rangle_{\rm fixed-shell})^2\rangle_{\rm fixed-shell}$ decreases with increasing shell thickness (see \ref{fig:figure6}c), reflecting how the shell hampers spatially-varying elastic deformation of the core. Comparison of \ref{fig:figure6}c and \ref{fig:figure4}b shows that $\langle \Delta {A}^2 \rangle_{\rm fixed-shell}$ also closely tracks changes in $\langle f_{\alpha} \rangle_{\rm fixed-shell}$. Local area distortions are thus strongly tied to local composition. 

\section*{Results in Three Dimensions}

\begin{figure*}
    \centering
    \includegraphics[width=17.8cm]{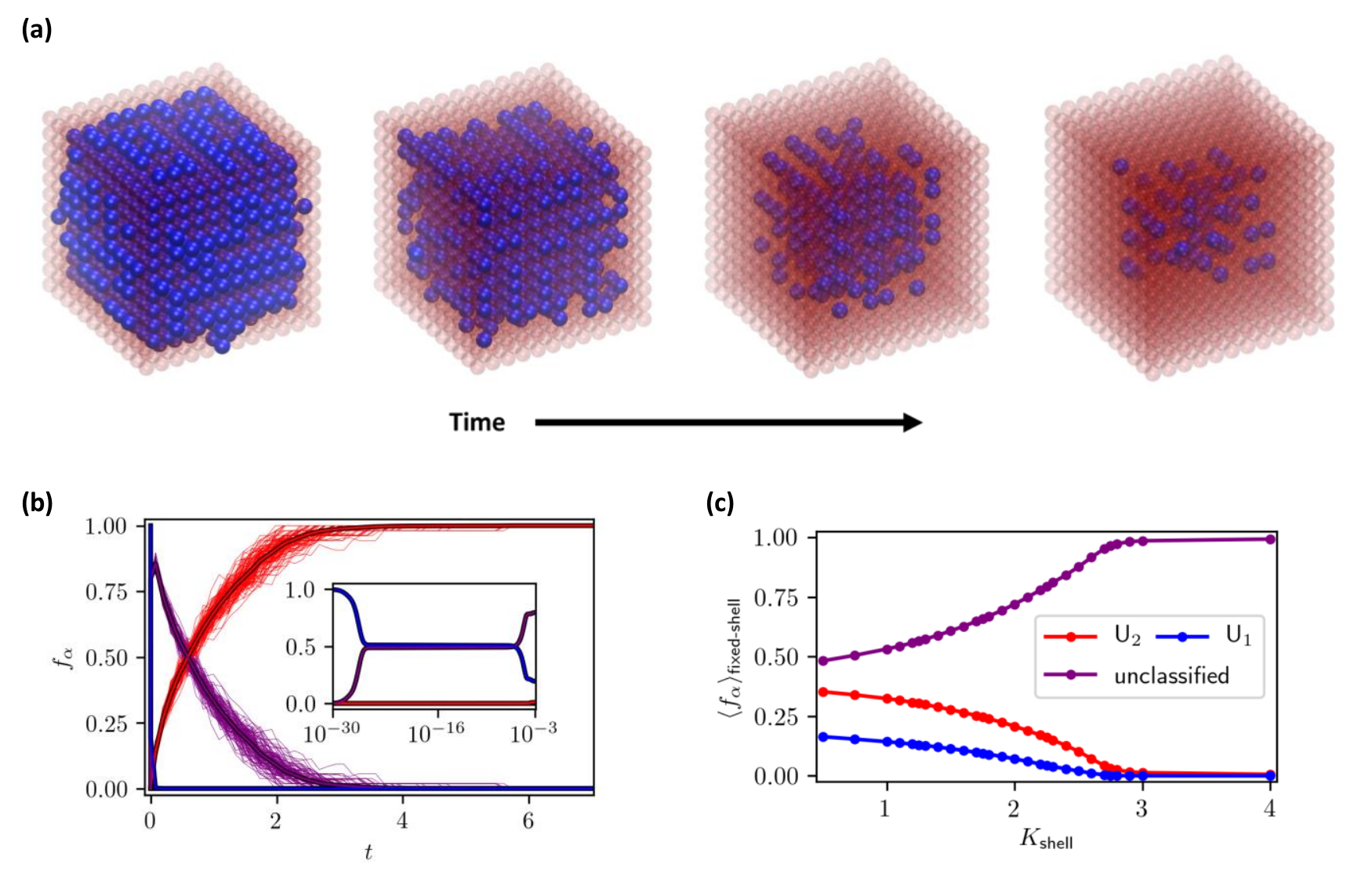}
    \caption{\textbf{(a)} Configurations taken from an elastic-model exchange trajectory of a 3d nanocrystal, whose atoms reside on a simple cubic lattice (with both nearest- and next-nearest-neighbor bonds). A atoms are colored blue; B atoms are colored red and are rendered as translucent for clarity. Note the spatially modulated patterns (columns of blue atoms) that develop over the course of time. \textbf{(b)} Fractional population $f_{\alpha}$ of $U_1$ and $U_2$ phases versus time; the unclassified population $1-f_{U}$ here includes contributions from spatially modulated phases as well as the compositionally disordered phase. Thin, solid-color lines represent individual KMC trajectories with $N=1728$; thick lines outlined in black represent averages over 100 trajectories. The inset shows the short-time behavior on a logarithmic scale. \textbf{(c)} Average fractional population of phases in equilibrium MC simulations of fixed-shell nanocrystals, as a function of shell stiffness $K_{\text{shell}}$ for a core size $N_{\text{core}}=1728$ and a fixed core composition $c=1/2$. The predominantly-A and predominantly-B phases ($U_1$ and $U_2$) are present for small values of $K_{\text{shell}}$ but are absent for large values.}
    \label{fig:figure7}
\end{figure*}

All the results we have presented thus far have been for two-dimensional systems (which have the advantage of being easily visualized). Nanocrystals in the laboratory are of course mostly three-dimensional entities. We therefore also performed simulations of a three-dimensional version of our elastic model (see \ref{fig:figure7}.) The results broadly agree with those in two dimensions: rapid surface exchange is followed by slower internal rearrangements (\ref{fig:figure7}b) characterized by spatially modulated patterns (\ref{fig:figure7}a). These patterns are more complicated than in 2d. Rather than characterizing them in detail, here we have simply identified unstructured regions ($U_1$, $U_2$) and labeled everything else as ``unclassified.'' As we show in \ref{fig:figure7}c, these unclassified regions dominate over unstructured phases at equilibrium in the presence of a sufficiently stiff shell.

\section*{Discussion}

Our model, while fairly simple, has proven able to reproduce some key features of cation exchange reactions. We have shown how a strongly favorable driving force for exchange creates, in effect, a nonequilibrium boundary condition, in which transport of material between the solution and the nanocrystal interior is shaped by the state of the perimeter. Nanocrystal morphologies during reaction -- observed to vary widely in experiments depending on cation identity -- result in our models from the interplay between such nonequilibrium boundary conditions and the nature of interactions between the exchanging species. Elastic interactions, in particular, lead to rich spatial patterns in the composition, akin to (somewhat larger-scale) modulated patterns observed in experiments \cite{robinson_spontaneous_2007,demchenko_formation_2008}. Although our work was inspired by a desire to understand cation exchange within nanocrystals, our present model could also apply to nanoparticle superlattices; in that context one could envision sites corresponding to individual nanocrystals, the length of whose DNA-mediated bonds could be tuned by pH, for example \cite{zhu_ph-responsive_2018}. Our results would then suggest altering the properties of the perimeter of such assemblies as a way of biasing spatial organization of the interior. Additionally, our finding that sufficiently stiff or large shells stabilize modulated phases in the nanocrystal core may explain the recent observation of ``labyrinth'' structures in a related elastic model of spin-crossover nanoparticles \cite{singh_elastic_2020}.

As a model of cation exchange in particular, our model clearly has shortcomings. It does not resolve anion and cation sublattices, and hence omits potentially important electrostatic interactions \cite{ott_microscopic_2014}. Additionally, it has fixed lattice connectivity and hence cannot describe dislocations or account explicitly for interstitials or vacancies One could imagine remedying this in part by introducing ``defect'' states (potentially carrying a net charge) in addition to the A and B states, or by allowing ``bonds'' between adjacent sites to break or form. Importantly, our model entirely ignores the ligands which passivate semiconductor nanocrystals and whose density on different facets could affect the rate of surface exchange. A more elaborate model might account for their effect implicitly -- perhaps by modifying the rates of surface exchange in the KMC scheme based on the facet and ligand type -- or might represent them explicitly as fluctuating species which modify $k_{\text{ex}}$ by their interactions with surface sites. In spite of these limitations, the success of our model in capturing the basic phenomenology of nanocrystal compositional change may serve as a foundation for future, more microscopically-detailed investigations.

\section*{Materials and Methods}

\subsection*{Energy function}

The total energy $E(C)$ of a configuration $C=\{\sigma_{\mathbf{R}}\}$ is given by:
\begin{equation}
    E(C)=E_{\text{elastic}}(C)+E_{\text{Ising}}(C)-(\Delta \mu/2)\sum_{\mathbf{R}}\sigma_{\mathbf{R}}.
\end{equation}
In simulations of the Ising model, we set the parameter $\epsilon$ in the elastic energy to zero; similarly, in simulations of the elastic model, we set the parameter $J$ in the Ising energy to zero.

In simulating the non-interacting model, we set $\Delta \mu/k_BT=-10$. For our Ising model simulations, we set $k_BT/J=2$ and $\Delta \mu/J=-10$. This temperature is well below the triangular-lattice Ising model critical temperature \cite{fisher_theory_1967}. Simulations of the 2d elastic model used $k_BT/\epsilon=0.2$ and $\Delta\mu/\epsilon=-20$.
For simulations of the 3d elastic model, we used $k_BT/\epsilon=1.5$ and $\Delta\mu/\epsilon=-200$. The choice of temperatures for elastic-model simulations would place our systems in the low-temperature regions of the corresponding bulk phase diagrams, where modulated phases are thermodynamically stable \cite{frechette_consequences_2019,frechette_chemical_2020}.

\subsection*{Elastic interaction potential}
The effective elastic potential $V_{\rm elastic}({\bf R},{\bf R}')$ is determined by integrating out the mechanical fluctuations of an elastic model which resolves atomic motions. To be concrete, we take the small-mismatch limit of that model's Hamiltonian \cite{frechette_consequences_2019,frechette_origin_2020}:
\begin{equation}
\begin{split}
    \mathcal{H}_{\text{elastic}}/\epsilon=2\sum_{\mathbf{R},\hat{\bm{\alpha}}(\mathbf{R})}&\Bigl(\hat{\bm{\alpha}}\cdot(\mathbf{u}_{\mathbf{R}}-\mathbf{u}_{\mathbf{R}+a\hat{\bm{\alpha}}})\Bigr.\\
    &\Bigl.-\frac{1}{2}(\delta \sigma_{\mathbf{R}}+\delta\sigma_{\mathbf{R}+a\hat{\bm{\alpha}}})-(\tilde{\sigma}_0/N-\delta a)\Bigr)^2,\label{eq:full_hamiltonian}
\end{split}
\end{equation}
where $\bm{\alpha}(\mathbf{R})$ are bond vectors which may vary by site (surface sites have fewer bonds than interior sites), $\tilde{\sigma}_0=\sum_{\mathbf{R}}\sigma_{\mathbf{R}}$, $\delta \sigma_{\mathbf{R}}=\sigma_{\mathbf{R}}-\tilde{\sigma}_0/N$, $a$ is the lattice parameter, and $\delta a = a-l_{AB}$ (where $l_{AB}=(l_{AA}+l_{BB})/2$, $l_{AA}$ and $l_{BB}$ are the rest lengths of A-A and B-B bonds respectively.) The displacement variables $\mathbf{u}_{\mathbf{R}}$ measure deviations of atoms from their perfect lattice sites $\mathbf{R}$. The energy scale $\epsilon=K\Delta^2/2$ depends on the spring constant $K$ of harmonic bonds between sites, as well as the lattice mismatch $\Delta=(l_{AA}-l_{BB})/2$.
Integrating over the harmonic mechanical fluctuations is equivalent to minimizing $\mathcal{H}$ with respect to $\delta a$ and $\mathbf{u}_{\mathbf{R}}$. The former simply contributes an additive constant to the free energy, which we ignore. The latter yields minimum-energy displacements for a given configuration of composition variables,
\begin{equation}
    \mathbf{u}_{\mathbf{R}}=\frac{1}{2}\sum_{\mathbf{R}'}\left(\mathbf{D}^{-1}\mathbf{C}\right)_{\mathbf{R},\mathbf{R}'}\sigma_{\mathbf{R}'}.\label{eq:min_energy_disp}
\end{equation}
The effective energy function $E_{\text{elastic}}$ that results from evaluating $\mathcal{H}_{\text{elastic}}$ at those displacements is:
\begin{equation}
E_{\text{elastic}}=\frac{1}{2}\sum_{\mathbf{R},\mathbf{R}'}\sigma_{\mathbf{R}}\left(\mathbf{S}-\frac{1}{4}\mathbf{C}^T\mathbf{D}^{-1}\mathbf{C}\right)_{\mathbf{R},\mathbf{R}'}\sigma_{\mathbf{R}'}
\end{equation}
where the matrices $\mathbf{D}$, $\mathbf{C}$, and $\mathbf{S}$ are given by:
\begin{align}
\mathbf{D}_{\mathbf{R},\mathbf{R}'}&=\sum_{\hat{\bm{\alpha}}(\mathbf{R})}\hat{\bm{\alpha}}(\delta_{\mathbf{R},\mathbf{R}'}-\delta_{\mathbf{R'},\mathbf{R}+a\hat{\bm{\alpha}}})\hat{\bm{\alpha}}\\
\mathbf{C}_{\mathbf{R},\mathbf{R}'}&=\sum_{\hat{\bm{\alpha}}(\mathbf{R})}\hat{\bm{\alpha}}(-\delta_{\mathbf{R},\mathbf{R}'}+\delta_{\mathbf{R'},\mathbf{R}+a\hat{\bm{\alpha}}})\\
\mathbf{S}_{\mathbf{R},\mathbf{R}'}&=\frac{1}{4}\sum_{\hat{\bm{\alpha}}(\mathbf{R})}(\delta_{\mathbf{R},\mathbf{R}'}+\delta_{\mathbf{R'},\mathbf{R}+a\hat{\bm{\alpha}}}).
\end{align}
We identify an effective pair potential as
\begin{equation}
V_{\rm elastic}({\bf R},{\bf R}')/\epsilon=\left(\mathbf{S}-\frac{1}{4}\mathbf{C}^T\mathbf{D}^{-1}\mathbf{C}\right)_{\mathbf{R},\mathbf{R}'}.
\end{equation} 
$V_{\rm elastic}$, which is sensitive to the nanocrystal's boundaries, was evaluated and tabulated for all site pairs $({\bf R},{\bf R}')$ in advance of each simulation.

\subsection*{Form of rate constants}

The rate of an event depends on the initial and final configurations $C$ and $C'$ through the energy function $E$. We chose rate constants
\begin{align}
k_{\text{ex}}&=k_{\text{ex}}^0e^{-\beta (E(C')-E(C))/2}\\
k_{\text{diff}}&=k_{\text{diff}}^0e^{-\beta (E(C')-E(C))/2}
\end{align}
that are consistent with the Boltzmann distribution, satisfying $k_{\text{ex}}(C\rightarrow C')/k_{\text{ex}}(C'\rightarrow C)=e^{-\beta (E(C')-E(C))}$ as a condition of detailed balance (and similarly for $k_{\text{diff}}$). The ``bare'' rate constants $k_{\text{ex}}^0$ and $k_{\text{diff}}^0$ set the fundamental timescales for exchange and diffusion. The mechanism of ion exchange depends only on their ratio, which we set as $k_{\text{ex}}^0/k_{\text{diff}}^0 = 1$ for simplicity. Surface dynamics are nonetheless much faster than diffusion for this case, as suggested by experiments \cite{groeneveld_tailoring_2013,bothe_solid-state_2015,kershaw_investigation_2017,moser_situ_2017}, due to the large chemical potential difference driving exchange. Accordingly, our results are qualitatively insensitive to the value of this ratio so long as $k_{\text{ex}}^0/k_{\text{diff}}^0\gg e^{\Delta \mu/(k_{\rm B} T)}$.

\subsection*{KMC algorithm}
 
KMC trajectories were generated using the Gillespie algorithm \cite{gillespie_exact_1977}. Specifically, given an initial configuration, we determine all possible exchange and diffusion moves and compute their rate constants. Denoting the rate constant of event $i$ (out of $n$ total) as $k_i$, we compute $k_{\text{total}}=\sum_{i=1}^n k_i$. We then generate a number $r$ uniformly at random from the interval $[0,k_{\text{total}}]$ and execute the event $j$ for which $\sum_{i=1}^{j-1}k_i\leq r < \sum_{i=1}^{j}k_i$. Finally, we update the time $t \rightarrow t+\Delta t$, where $\Delta t$ is sampled from the exponential distribution $P(\Delta t)=k_{\text{total}}\exp{(-k_{\text{total}}\Delta t)}$. Repeating this process many times yields a single trajectory.

\subsection*{Order parameter}

Consider a site $\mathbf{R}$ with nearest neighbors $\{\mathbf{N}\}$, and denote the mutual nearest neighbors between $\mathbf{R}$ and $\mathbf{N}$ as $\{\mathbf{N}'\}$. The order parameter $s$ associated with $\mathbf{R}$ is defined as:

\begin{equation}
    s(\mathbf{R})=
    \begin{cases}
    \begin{aligned} -2&, &&\text{if $\sigma_{\mathbf{R}}=\sigma_{\mathbf{N}}=-1$ for all $\mathbf{N}$}\\
    -1&, &&\parbox[t]{1.0\textwidth}{if $\sigma_{\mathbf{R}}=1$ and $\sigma_{\mathbf{N}}=-1$ for all $\mathbf{N}$, \\ OR $\sigma_{\mathbf{R}}=-1$ and $\sigma_{\mathbf{N}}=-\sigma_{\mathbf{N'}}$ for all $\mathbf{N}$, $\mathbf{N}'$}\\
    1&, &&\parbox[t]{1.0\textwidth}{if $\sigma_{\mathbf{R}}=-1$ and $\sigma_{\mathbf{N}}=1$ for all $\mathbf{N}$,\\  OR $\sigma_{\mathbf{R}}=1$ and $\sigma_{\mathbf{N}}=-\sigma_{\mathbf{N'}}$ for all $\mathbf{N}$, $\mathbf{N}'$}\\
    2&, &&\text{if $\sigma_{\mathbf{R}}=\sigma_{\mathbf{N}}=1$ for all $\mathbf{N}$}\\
    0&, &&\text{otherwise.}
    \end{aligned} 
    \end{cases} \label{eq:order_parameter}
\end{equation}

\subsection*{Calculation of local area deviations}

In the small-mismatch limit to which we restrict ourselves, the minimum-energy displacements $\{\mathbf{u}_{\mathbf{R}}\}$ are uniquely determined by the compositions $\{\sigma_{\mathbf{R}}\}$ according to \ref{eq:min_energy_disp}.
From the displacements, we can measure local deviations from the average crystal area (in 2D) due to heterogeneous composition configurations. Specifically, we compute the local area deviation associated with site $\mathbf{R}$ as:
\begin{equation}
    \Delta A_{\mathbf{R}}=\frac{1}{Za^2}\sum_{\hat{\bm{\alpha}}(\mathbf{R})}|\mathbf{u}_{\mathbf{R}+a\hat{\bm{\alpha}}}-\mathbf{u}_{\mathbf{R}}+a\hat{\bm{\alpha}}|^2-1,
\end{equation}
expressed in units of $a^2$. Here, $Z$ is the coordination number of the lattice, which is 6 for all calculations presented.

\subsection*{Equilibrium MC simulations of fixed-shell nanoparticles}

Equilibrium configurations were generated at fixed net composition using Kawasaki dynamics \cite{kawasaki_diffusion_1966}. Proposed moves consist of attempts to swap the identities $\sigma_{\mathbf{R}}$, $\sigma_{\mathbf{R}'}$ of two randomly selected sites $\mathbf{R}$, $\mathbf{R}'$ within the nanocrystal core. Such moves are accepted with probability \cite{frenkel_understanding_2002}:
\begin{equation}
P(C\rightarrow C')=\min\left[1,e^{-\beta(E(C)-E(C'))}\right].
\end{equation}
A single MC sweep consists of $N_{\text{core}}$ attempted moves. 

In each equilibrium fixed-shell simulation, core sites were initialized with random identities consistent with a net core composition of $c=1/2$. Systems were then equilibrated by running $10^{5}$ MC sweeps without collecting data. After the equilibration period, configurations were recorded every sweep for $10^5$ sweeps. Reported observables were averaged over these configurations.

Stiff-shell nanocrystals were modeled by changing the spring constant of the bonds connecting shell atoms; explicitly, terms in \ref{eq:full_hamiltonian} corresponding to shell-shell bonds were multiplied by a factor $K_{\text{shell}}/K$ before the energy minimization described previously to obtain $V_{\rm elastic}({\bf R},{\bf R}')$.

To mitigate potentially large surface contributions to quantities like $f_\alpha$, equilibrium observables were spatially averaged only over interior core sites -- the outermost core atoms were excluded.

\section*{Acknowledgments}

This work was supported by National Science Foundation (NSF) grant CHE-1416161. This research also used resources of the National Energy Research Scientific Computing Center (NERSC), a U.S. Department of Energy Office of Science User Facility operated under Contract No. DE-AC02-05CH11231.  P.L.G. and L.B.F. acknowledge stays at the Erwin Schr{\"o}dinger Institute for Mathematics and Physics at the University of Vienna.

\FloatBarrier

\bibliography{ref.bib}

\end{document}